\documentclass{PoS}

\title{Dark Matter search with H.E.S.S. towards ultra-faint dwarf nearby DES satellites of the Milky Way}

\ShortTitle{DM search with H.E.S.S. towards ultra-faint DES satellites of the Milky Way}

\author{\speaker{Lucia Rinchiuso}\\
        IRFU, CEA, Universit\'e Paris-Saclay, F-91191 Gif-sur-Yvette, France\\
        E-mail: \email{lucia.rinchiuso@cea.fr}}

\author{Emmanuel Moulin\\
        IRFU, CEA, Universit\'e Paris-Saclay, F-91191 Gif-sur-Yvette, France\\
        E-mail: \email{emmanuel.moulin@cea.fr}}

\author{C\'eline Armand\\
        Laboratoire d'Annecy-le-Vieux de Physique des Particules, Universit\'e Savoie Mont-Blanc, CNRS/IN2P3, F-74941 Annecy-le-Vieux, France\\
        E-mail: \email{celine.armand@lapp.in2p3.fr}}

\author{Vincent Poireau\\
        Laboratoire d'Annecy-le-Vieux de Physique des Particules, Universit\'e Savoie Mont-Blanc, CNRS/IN2P3, F-74941 Annecy-le-Vieux, France\\
        E-mail: \email{poireau@lapp.in2p3.fr}}       

\author{for the H.E.S.S. Collaboration\footnote{For collaboration list see PoS(ICRC2019)1177}}

\abstract{Several nearby ultra-faint satellites of the Milky Way discovered by the Dark Energy Survey (DES) during the last few years are promising targets for indirect dark matter (DM) searches with very-high-energy (VHE, E$\gtrsim$100 GeV) gamma rays. The H.E.S.S. experiment has carried out an observational campaign in 2017 and 2018 towards a selection of the most promising DES dwarf satellites, accumulating a total observation time of about 80 hours. The individual datasets have been used to look for a DM signal in several annihilation channels. No significant VHE gamma-ray excess above the background is found in any of the targets. Constraints are derived on the thermally-averaged velocity-weighted annihilation cross section $\langle \sigma v \rangle$ of the DM particles versus their mass. A combined analysis of the datasets has been performed. The strongest 95\% C.L. upper limits reach a $\langle \sigma v \rangle$ value of a few  $10^{-24}$~cm$^3$~s$^{-1}$ for the DM continuum signals in the TeV DM mass range. They are the first constraints derived from imaging atmospheric Cherenkov telescope observations towards DES dwarf galaxy satellites. These limits are among the most constraining so far in the TeV DM mass range towards dwarf satellites of the Milky Way.}

\FullConference{36th International Cosmic Ray Conference -ICRC2019-\\
		July 24th - August 1st, 2019\\
		Madison, WI, U.S.A.}

\begin{document}

\section{Introduction}
The presence of dark matter (DM) in the Universe is needed at all scales from a wealth of cosmological and astrophysical measurements. However, its underlying nature remains unknown. A leading class of elementary particle candidates is made of weakly interacting massive particles (WIMP). Provided they are produced thermally in the early Universe, their mass and coupling strength at the electroweak scale make their relic density consistent with that of observed DM.

WIMPs can be searched for indirectly through the detection of the final state products of their self-annihilation such as gamma rays. Arrays of Imaging Atmospheric Cherenkov Telescopes (IACTs) have been built to detect very-high-energy ($\gtrsim100$~GeV) gamma rays.
Due to their relatively small field of view ($\sim5^\circ$) the IACT observation strategy for dark matter detection is focused on regions of the sky with the largest expected DM density and at a relatively short distance from Earth. 
Among them are the Galactic centre region and nearby dwarf galaxy satellites of the Milky Way.

Dwarf spheroidal galaxies (dSphs) are the most DM-dominated systems in the universe. While the expected DM signals are smaller than that of the Galactic Centre region, its detection in these systems could be unambiguous given the absence of  detection of conventional astrophysical emissions so far. 
Numerous dSphs are hosted in the Local Group and recent optical surveys such as DES~\cite{DES} revealed a new population of ultra-faint systems. Several of them can be well observed from the Southern hemisphere. The H.E.S.S. (High Energy Spectroscopic System) array  located in Namibia is particularly well suited for such observations. The H.E.S.S. instrument is composed of four 12-m diameter telescopes and a 28-m diameter telescope in the middle of the array. The five telescopes can work in combination and exploit the spectroscopy for a better reconstruction of the detected event and its position. H.E.S.S. reaches an energy resolution of $\Delta E/E=10\%$, and an angular resolution better than $0.1^\circ$ per gamma ray, which provides great opportunities in measuring the energy spectrum and the morphology of possible signals in dSphs.

The energy-differential flux in gamma rays  in the solid angle $\Delta\Omega$ expected from the self-annihilation of DM particles of mass $m_{\rm DM}$ writes:
\begin{equation}
\label{eq:flux}
\frac{{\rm d} \Phi_\gamma}{{\rm d} E_\gamma} (E_\gamma,\Delta\Omega)=
\frac {\langle \sigma v \rangle}{8\pi m_{\rm DM}^2}\sum_f  BR _f \frac{{\rm d} N^f}{{\rm d}E_\gamma} \, J(\Delta\Omega) \ ,
\quad {\rm with} \quad  J(\Delta\Omega) =  \int_{\Delta\Omega} \int_{LOS}\rho^2(s(r,\theta)) ds\, d\Omega \, .
\end{equation}
 $\langle \sigma v \rangle$ refers to as the total velocity-weighted annihilation cross section and $dN^f/dE_\gamma$ 
is the annihilation spectrum in the final state $f$ with associated branching ratio $BR_f$. The term $J$, referred to as the {\it J-factor}, corresponds to the integration of the square of the DM mass density $\rho$ over the line-of-sight (LOS) $s$ and $\Delta\Omega$. The distance from the observer to the annihilation location $s$ is given by $r = (s^2+r_0^2 - 2\,r_0\, s\,cos \theta)^{1/2}$, where $r_0$ is the distance from the target to the Earth and $\theta$ the angle between the direction of observation and the dSph centre. The DM density distribution can be inferred via the Jeans equation from the measurements of the position and line-of-sight velocity of the stars that are gravitationally bound in the dwarf galaxy potential well. A statistical uncertainty on the J-factor is obtained to the finite number of kinematic measurements of the member stars~\cite{Geringer-Sameth:2014yza}.

\section{Selection of nearby DES dwarf galaxy satellites of the Milky Way for observations with H.E.S.S.}
The Dark Energy Survey (DES) has recently discovered new ultra-faint Milky Way satellite systems that are candidates for dSphs. 
Due to their proximity and the large expected DM content, the DES dSph candidates are promising targets for DM searches. From their stellar dynamics and, when available, accurate spectroscopic measurements, J-factors as large as $\rm log_{10} (J/GeV^2cm^{-5}) \sim 19$ have been derived. 

Several DES dSphs with the large J-factor are located in the Southern hemisphere and therefore can be observed under favorable conditions from the H.E.S.S. site. 
H.E.S.S. has performed a two-year observation campaign towards a selection of DES dSphs.
The targets are selected as being the closest, with the largest expected DM content, and good visibility window. They are Reticulum~II (Ret~II), Tucana~II (Tuc~II), Tucana~III (Tuc~III),  and Grus~II (Gru~II). Tucana~IV (Tuc~IV) is also  observed as it is in the field of view of Tuc~III observations. The overall observation campaign on the selected targets amount to $\sim$80~hours.

The J-factors and their statistical uncertainty in a region of $0.5^\circ$ are provided in Tab.~\ref{tab:table1}.  
The J-factors for Tuc~III, Tuc~IV and Gru~II are obtained from predictions since high-quality spectroscopic
 measurements are not available yet for these objects. 
\begin{table}
\centering
\begin{tabular}{ c |  c | c | c  }
\hline
\hline
Source  &$\overline{\log_{10} J_{<0.5^{\circ}}}$ & $\sigma_{\rm J_{<0.5^{\circ}}}$ &   Ref.\\
name & [$\log_{10} {\rm (GeV^{5}cm^{-2}})$] &  [$\log_{10} {\rm (GeV^{5}cm^{-2}})$]  &\\
\hline
Reticulum II & 19.6 & 0.85 &  \cite{Bonnivard:2015tta}\\
Tucana II & 18.7 & 0.80 & \cite{Walker:2016mcs}\\
Tucana III* & 19.4 & - &  \cite{Simon:2016mkr}\\
Tucana IV* & 18.7  & - & \cite{Fermi-LAT:2016uux}\\
Grus II* & 18.7 & - & \cite{Fermi-LAT:2016uux}\\
\hline
\hline
\end{tabular}
\caption{\label{tab:table1}
Selected Milky Way satellites discovered by DES for H.E.S.S. observations. The second and third columns provide the J-factor mean value for an integration angle of 0.5$^{\circ}$ and its associated 1$\sigma$ statistical uncertainty. The references from which these values are extracted are given in the fourth column. The symbol * marks the systems for which no spectroscopic measurement is available.}
\end{table}

\section{Data analysis and search for dark matter signals}
The observations on the selected DES dSphs are carried out with the array including the five telescopes in the data taking. A gamma-ray signal from DM annihilation is looked for over a background which consists of misidentified hadrons, commonly referred to as the residual background. The DM signal is searched in a region of interest, the {\it ON} region, and the residual background in control regions called as the {\it OFF} regions. The ON region is defined as a disk centered at the nominal position of the target.  The disk radius is $0.2^\circ$ for the targets with measured J-factors, which 
enable to contain more than 97\% of the expected DM signals. For dSphs that lack of accurate spectroscopic measurements, they are treated as point-like objects with disk radii of $0.125^\circ$.  

In case of extended sources the ON region is divided in two sub-regions of interest (ROIs) of width $0.1^\circ$ each.
Multiple OFF regions are defined for each ROI as regions with the same shape and solid angle as the ROI. The OFF regions lie on a ring centered at the telescope pointing position, at the same distance as the ON region from the  pointing position of the telescopes. An exclusion region disk of radius twice the ON region radius is used in order to avoid any possible signal contamination in the OFF regions. An example of the background determination procedure is shown in Fig.{fig:Bckdetemination} in the case of Ret~II.
\begin{figure}
\centering
\includegraphics[scale=0.5]{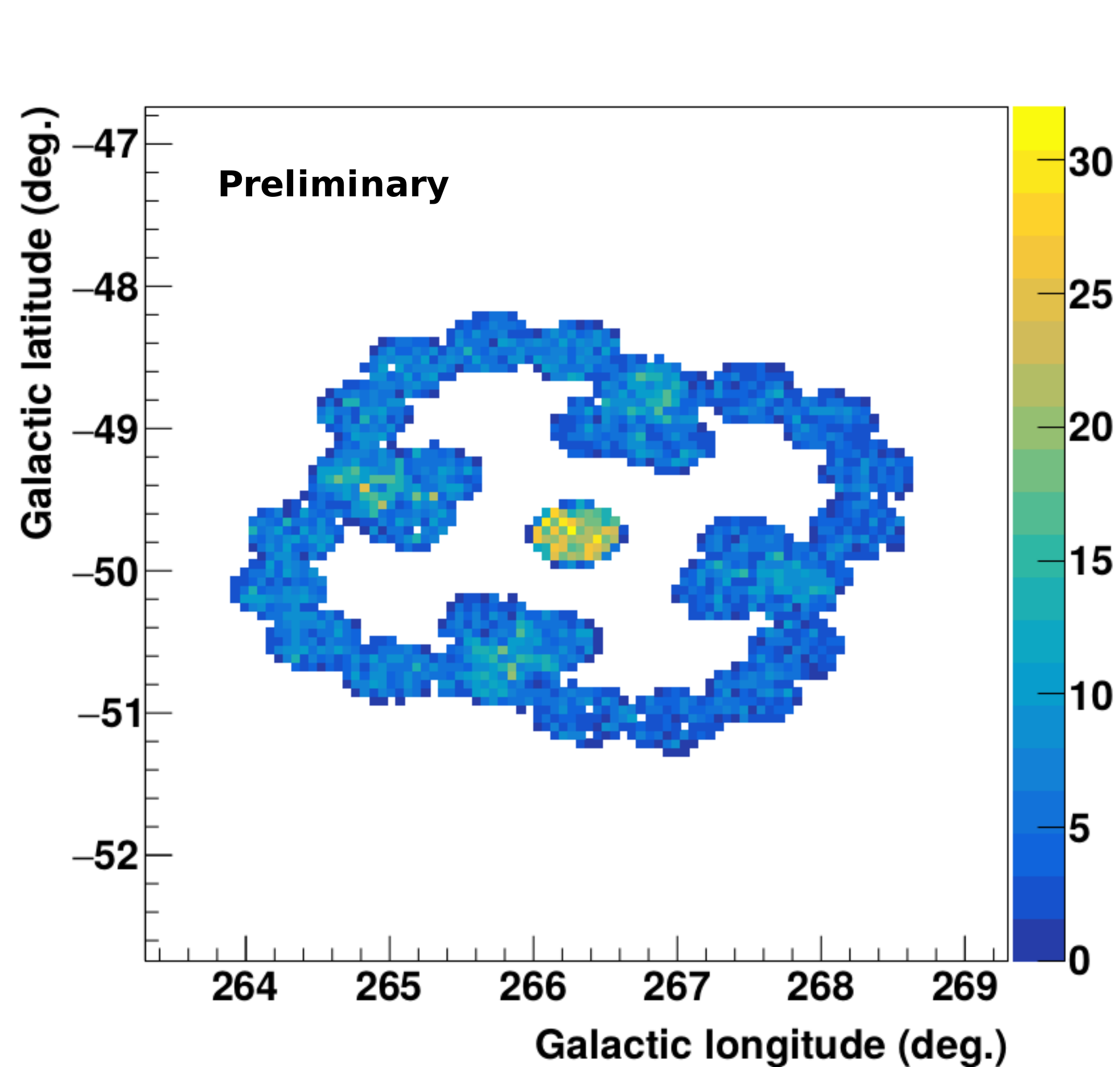}
\caption{Residual background determination using the Multiple OFF method for the ultra-faint dwarf galaxy Ret II. The background is measured in OFF regions taken in a ring centered at the telescope pointing position, at the same distance as the ON region from the  pointing position of the telescopes. An exclusion region is taken around the ON source region as a disk of radius twice the ON region radius. The color scale provides the number of event count in pixels of size 0.02$^\circ \times$0.02$^\circ$.}
\label{fig:Bckdetemination}
\end{figure}

\begin{figure}
\centering
\includegraphics[scale=0.5]{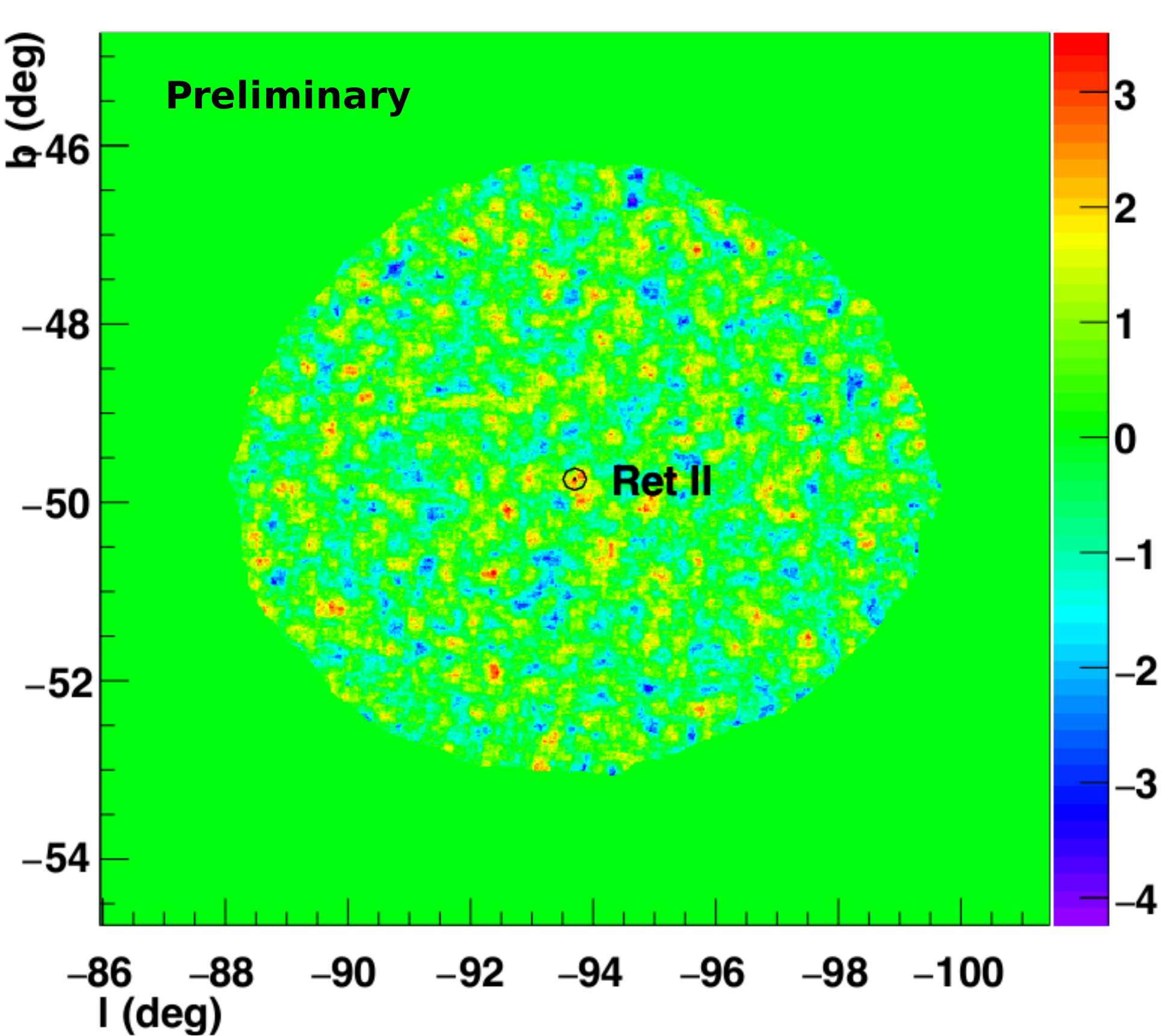}
\caption{Excess significance map in the field of view of Ret II in Galactic coordinates. The nominal position is marked by a black circle. The color scale provides the significance of the excess count in units of $\sigma$ in pixels of size 0.02$^\circ \times$0.02$^\circ$. No significant excess is found in the field of view.}
\label{fig:SignificanceMap_RetII}
\end{figure}
The number of measured events in the ON and OFF regions, $N_{\rm ON}$ and $N_{\rm OFF}$, respectively, are computed. The ratio between the solid angle size of the OFF and ON regions is defined as $\alpha$. $N_{\rm ON}$ and $N_{\rm OFF}/\alpha$ are compared for each target. Fig.~\ref{fig:SignificanceMap_RetII} shows the excess significance map in Galactic coordinates in the filed of view of Ret~II.
No significant excess above the background is found in Ret~II as well as in any of other targets.

In order to derive constraints on the DM properties in each system, a  2  dimensional  (2D)-binned  Poisson  maximum  likelihood  analysis  is  performed  to  exploit  the  spatial  and spectral features of the expected DM signal with respect to the background. For a given DM mass and annihilation channel, the Poisson likelihood function writes in the energy bin $i$ and spatial bin $j$ as a product of ON et OFF Poisson distributions, with $N_{\rm ij}^{\rm S} + N_{\rm ij}^{\rm B}$ and  $\alpha N_{\rm ij}^{\rm B}$ the expected values in the ON and OFF regions, respectively. The expected DM signal in the bin $(i,j)$, $N_{\rm ij}^{\rm S}$, is obtained by folding the theoretical DM  flux given in Eq.~(\ref{eq:flux})  with  the  energy-dependent acceptance and energy resolution of H.E.S.S. for the considered data set. 

Since no significant excess is found in any of the system, upper limits can be derived for any DM mass from a log-likelihood ratio test statistic  TS following the procedure described in Ref.~\cite{Cowan:2010js}.
A value of TS = 2.71 corresponds to one-sided upper limits on $\langle\sigma v\rangle$ at a 95\% confidence level. The expected limits are derived by performing Poisson realizations of the measured counts $N_{\rm OFF}$ and $N_{\rm OFF}/\alpha$ and computing the TS value for each realization.
The mean of the distribution of the values of $\langle\sigma v\rangle$ corresponds to the mean expected limits. The standard deviation of the $\langle\sigma v\rangle$ distribution provides the $1\sigma$ containment band of the expected limits.

\section{Results}
\begin{figure}
\centering
\includegraphics[scale=0.5]{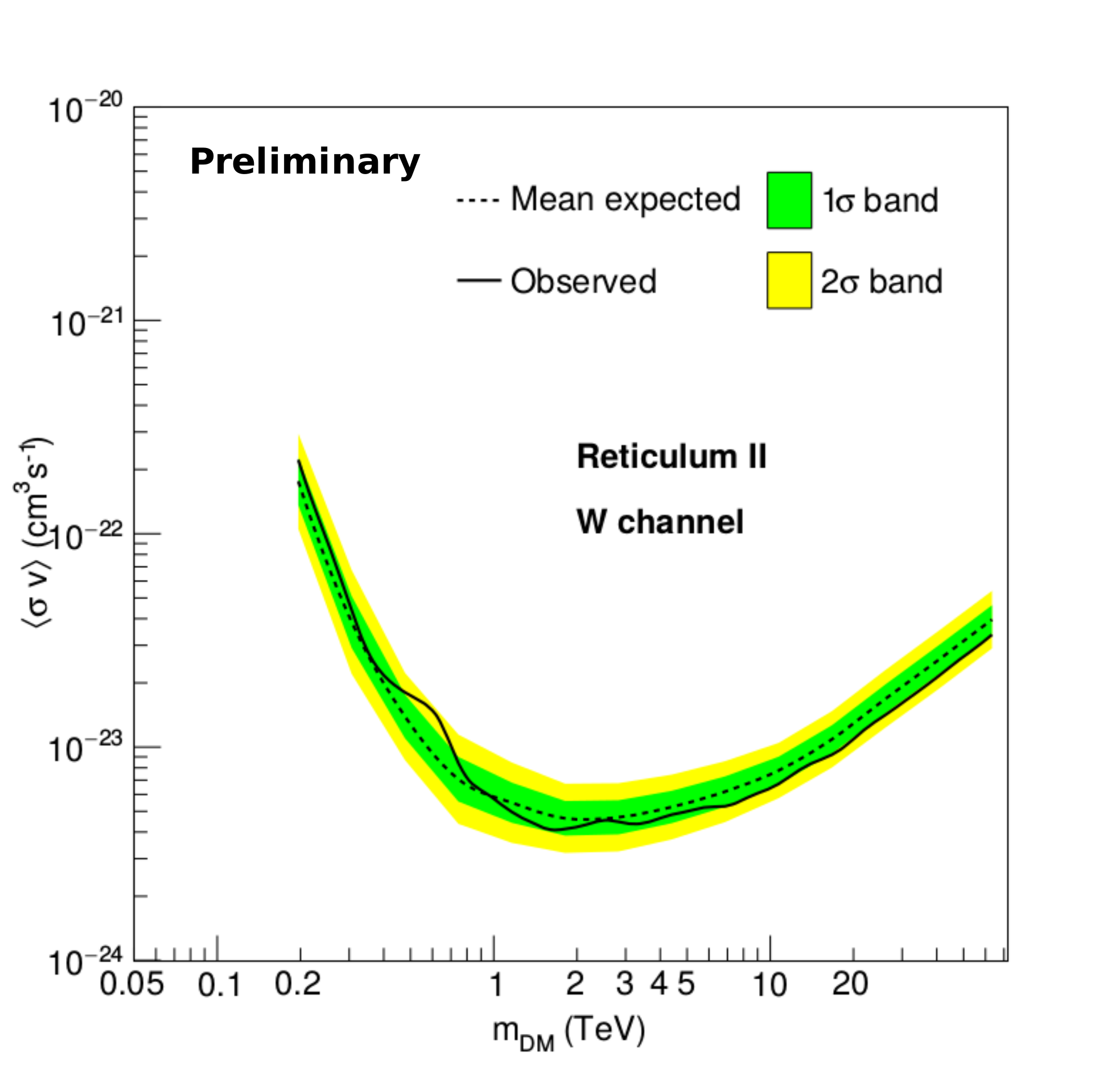}
\caption{95\% C.L. upper limits on the annihilation cross section $\langle\sigma v\rangle$ for Ret~II in the $W^+W^-$ annihilation channel. Observed limits (solid lines) together with mean expected limits (dashed line) and the 1$\sigma$ (green area) and 2$\sigma$ (yellow area) containment bands are shown, respectively. No statistical uncertainty on the J factor is taken into account here.}
\label{fig:Limits_W_RetII}
\end{figure}

Constraints on $\langle\sigma v\rangle$ as function of the DM mass are computed in terms of 95\% C.L. upper limits 
for each target and for several annihilation channels. Fig.~\ref{fig:Limits_W_RetII} shows the 95\% C.L. upper limits on the annihilation cross section $\langle\sigma v\rangle$ for Ret~II in the $W^+W^-$ annihilation channel. The observed limit reaches 6$\times$10$^{-24}$ cm$^3$s$^{-1}$ for a DM particle mass of 1 TeV.

\begin{figure}
\centering
\includegraphics[scale=0.5]{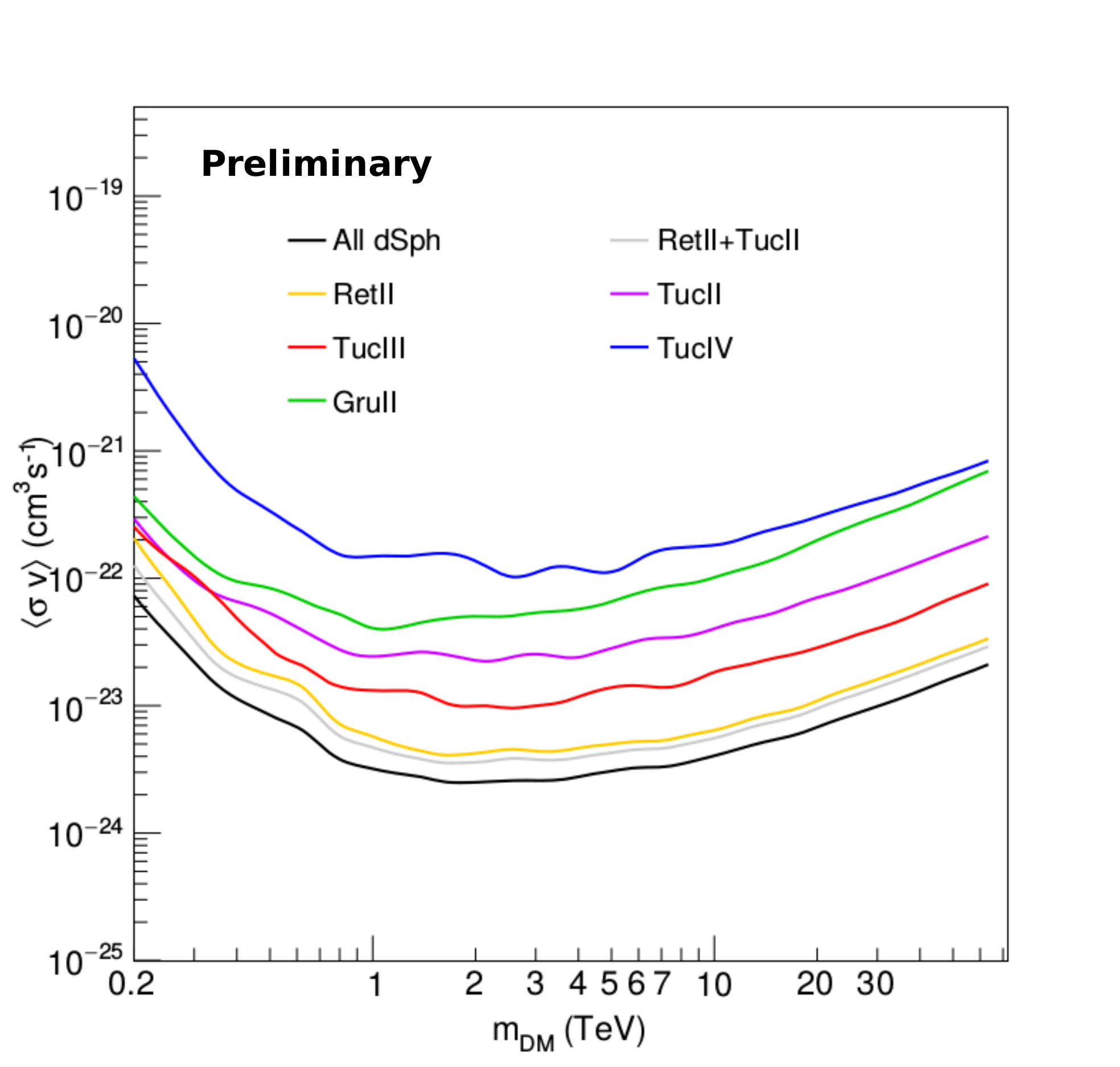}
\caption{95\% C.L.  observed upper limits on the annihilation cross section $\langle\sigma v\rangle$ for the individual and combined datasets of the target in the $W^+W^-$ annihilation channel. No statistical uncertainty on the J factor is taken into account here.}
\label{fig:Limits_W_all}
\end{figure}

In order to improve the H.E.S.S. sensitivity to a DM annihilation signal the observational datasets on the targets can be combined at the likelihood level and the test statistics can be performed on the combined likelihood over the targets. No significant excess is found in the combined analysis of the datasets.  Fig.~\ref{fig:Limits_W_all} shows the 95\% C.L.  observed upper limits on the annihilation cross section $\langle\sigma v\rangle$ for the individual and combined datasets  in the $W^+W^-$ annihilation channel. The combined observed limit is mostly driven by the Ret~II and Tuc~III contributions. The observed combined limit reaches 3$\times$10$^{-24}$ cm$^3$s$^{-1}$ for a DM particle mass of 1 TeV.

\section{Summary}
The H.E.S.S. observation carried out a fruitful observation campagain towards a selection of recently-detected ultra-faint dwarf galaxy satellites of the Milky Way recently discovered by DES to search for a DM annihilation signal in VHE gamma rays. The observations towards the five selected targets amount to about 80 hours of live time in total. 
In absence of any significant excess in any field of views, 95\% C.L. upper limits have been derived on the velocity-weighted annihilation cross section of DM particles as a function of their mass in various annihilation channels. The strongest results on an individual object are obtained for Reticulum II.  
The new results obtained by the H.E.S.S. collaboration are particularly relevant in the context of TeV DM models. They nicely complement the Fermi-LAT limits obtained at lower DM masses. 

\section*{Acknowledgements}
Full H.E.S.S. acknowledgements for contributions to conferences in 2019 can be found here: 
\href{https://www.mpi-hd.mpg.de/hfm/HESS/pages/publications/auxiliary/HESS-Acknowledgements-2019.html}{https://www.mpi-hd.mpg.de/hfm/HESS/pages/publications/auxiliary/HESS-Acknowledgements-2019.html}.

\end{document}